\documentclass[12pt]{article}
\usepackage{natbib,graphicx,setspace,amsmath,amssymb,subfigure}
\newcommand{\boldmu}{\boldsymbol{\mu}}
\begin{document}

\title{Sparse Bayesian Hierarchical Modeling of High-dimensional Clustering Problems}         
\author{Heng Lian\\Division of Mathematical Sciences\\School of Physical and Mathematical Sciences\\Singapore 637371\\Singapore}        
\date{}          
\maketitle
\begin{abstract}
Clustering is one of the most widely used procedures in the analysis of microarray data, for example with the goal of discovering cancer subtypes based on observed heterogeneity of genetic marks between different tissues. It is well-known that in such high-dimensional settings, the existence of many noise variables can overwhelm the few signals embedded in the high-dimensional space. We propose a novel Bayesian approach based on Dirichlet process with a sparsity prior that simultaneous performs variable selection and clustering, and also discover variables that only distinguish a subset of the cluster components. Unlike previous Bayesian formulations, we use Dirichlet process (DP) for both clustering of samples as well as for regularizing the high-dimensional mean/variance structure. To solve the computational challenge brought by this double usage of DP, we propose to make use of a sequential sampling scheme embedded within Markov chain Monte Carlo (MCMC) updates to improve the naive implementation of existing algorithms for DP mixture models. Our method is demonstrated on a simulation study and illustrated with the leukemia gene expression dataset.  

\textbf{keywords:} Dirichlet process; Markov chain Monte Carlo; Sequential sampling; Sparsity prior; 
\end{abstract}

\section{Introduction}       
Clustering is one of the most widely used procedures in the analysis of microarray data. It has been used, for example, for cancer subtype discovery \citep{golub99}. Technological advances over the last decade on microarrays have made possible simultaneous investigation of thousands of genes that potentially characterize and distinguish previously known or unknown cancer subtypes. Although obviously not all the genes arrayed possess discriminative power for different cancer subtypes, if fewer genes are used, the procedure might fail to distinguish between some of the subtypes. Also, because of the cost of arraying the transcripts, this is a typical ``large p, small n" problem that has attracted much attention recently. 

Among many classes of clustering procedures, the model-based approach \citep{banfield93, fraley02}, assuming the data come from a mixture of distributions, has the advantage of permitting principled statistical inferences compared to other procedures based largely on heuristics, such as k-means. This is especially important in our case where inferences should be made on the selected variables as well as clustering structure. 

Motivated by model interpretation as well as parsimony considerations, variable selection in clustering, mostly within the Bayesian framework, has been of increasing interest. Compared to variable selection in regression \citep{tibshirani96, george93, george97}, the clustering problem is much less studied. \cite{friedmanjrss04} proposed one approach to select different subsets of variables and different associated weights for different clusters for non-model-based clustering. \cite{liu03} proposed to first reduce dimension by performing the principal component analysis on the covariates and then fitting a Bayesian mixture model to the top factors. Although the number of factors is automatically determined by the model, there are several disadvantages to this approach, including difficulty in interpretation in terms of the original attributes. Also, it can be argued that the top principal components do not necessarily have the most significant discriminative power for clustering and thus the procedure is suboptimal. \cite{tadesse05} adapted the stochastic search strategy of \cite{george93,george97} originally proposed in the regression context and used reversible jump MCMC for inferences of cluster structures with simultaneous variable selection. This approach assumes that the same subsets of covariates discriminate all clusters. The model laid out in \cite{kim06} is based on the same philosophy but utilized an infinite mixture model via the Dirichlet process (DP) mixture. On the other hand, \cite{hoff06} adopted a mean shift approach in which each cluster-specific mean deviates from the global base-line mean on one or more attributes that differs from cluster to cluster. 

In this paper, we propose a Bayesian model for simultaneous clustering and variable selection via DP mixture as well. Our formulation is based on the mean shift model \citep{hoff06}. However, we use a novel hierarchical sparsity prior similar to that of \cite{lukas06} which can improve separation of significant signals from noise variables and thus can lead to reduced false discoveries. Also, we use a Dirichlet process shrinkage approach for both high-dimensional mean and variance that outperforms shrinkage using a non-DP prior, typically with normal distribution for mean and inverse-Gamma distribution for variance. Because of this double usage of Dirichlet process, both for sample clustering and for covariate shrinkage, the direct implementation of standard DP algorithms available in the literature becomes very inefficient. We solve this problem by utilizing an embedded sequential sampling step as the proposal distribution in the Markov chain Monte Carlo (MCMC) iterations. In the next section, we formulate our model using the sparsity prior. Posterior computation via MCMC is discussed in detail in Section 3 where we also show how to use sequential sampling for efficient updating. For clarity in exposition, these two sections only consider shifts in means. Extension to shifts in both means and variances is briefly considered in Section 4. Section 5 includes a simulation study as well as an application to the leukemia gene expression data. We conclude the article with a brief discussion in Section 6. 

\section{Model Formulation}    
In this section as well as the next, we consider the case where the clusters differ from each other only in terms of their respective means for some of the attributes. In our model we start by expressing the samples $\mathbf{y}_i=(y_{i1},\ldots,y_{ip}), i=1,\ldots, n$, as
\[y_{ij}=\mu_j+\mu_{ij}+\sigma_j\epsilon_{ij},\;\epsilon_{ij}\stackrel{i.i.d.}{\sim}N(0,1).\]
In this formulation, $\mu_j$ and $\sigma_j$ are attribute-specific mean and standard deviations shared by all samples. We put the following priors for them:
\[\mu_j\stackrel{i.i.d.}{\sim}DP(\alpha N(\mu_0,\sigma_0)),\]
\[\sigma_j^2\stackrel{i.i.d}{\sim}DP(\beta Inv-Gamma(\alpha_0,\beta_0)),\]
where $DP(\alpha H)$ is the Dirichlet process with concentration parameter $\alpha>0$ and base probability measure $H$. In this paper, we use the notation $\theta_i\stackrel{i.i.d.}{\sim} DP(\alpha H)$ as a short form for the more rigorous $\theta_i\stackrel{i.i.d.}{\sim}G, G\sim DP(\alpha H)$. This might be a misuse but simplifies our notation since DP appears multiple times at different places within our model. When $\alpha\rightarrow \infty$, the first expression above reduces to $\mu_j\sim N(\mu_0,\sigma_0)$, for example. The use of Dirichlet process can be motivated from at least two point of views. First, it relaxes the normality assumption imposed on the components of the mean vector. Second, since the DP is a discrete measure, it provides a regularization mechanism by shrinking different parameters towards each other. 

Since the attribute specific $\mu_j$ and $\sigma_j$ are shared by all samples, the clustering structure can only derive from appropriate specification on $\mu_{ij}$. As in \cite{hoff06,kim06}, the clustering of samples will be determined by an infinite mixture of distributions via Dirichlet process mixture. Denote ${\boldmu}_i=(\mu_{i1},\ldots,\mu_{ip})$. When it is intended that $\boldmu_i$ is the mean for cluster $c$, i.e. sample $i$ is assigned to cluster $c$, we also use $\boldmu_c$ to denote the same mean vector. Although there might be some concern over misuse of notation, this can hardly cause any confusion in the context. The sample means are generated from an infinite mixture specified as the following:
\[\boldmu_i=(\mu_{i1},\ldots,\mu_{ip})\stackrel{i.i.d.}{\sim}DP(\tau H),\]
where the base measure $H$ on $\boldmu_i$ can be defined through the following hierarchical ``point-mass mixture" prior:
\begin{eqnarray*}
\mu_{ij}&\sim& (1-\pi_{ij})\delta_0+\pi_{ij}DP(\gamma N(0,\eta_i^2)),\\
\pi_{ij}&\sim& (1-\rho_j)\delta_0+\rho_j Beta(a,b),\\
\rho_j&\sim&Beta(c,d).
\end{eqnarray*}
Thus in our model, not only are samples assigned to different groups (i.e., $\boldmu_i=\boldmu_j, 1\le i\neq j\le n$ with positive probability), the nonzero components of the mean specific to a cluster are also clustered (i.e., $\mu_{ij}=\mu_{ik}, 1\le j\neq k\le p$ with positive probability). In this paper, we choose to use a more parsimonious model $\eta_i\equiv\eta$. The prior structure presented above has individual probability $\pi_{ij}$ that attribute $j$ has a nonzero effect for cluster $c$ to which the $i$-th sample is assigned, while the attribute specific parameter $\rho_j$ indicates the sparsity propensity of the covariate $j$. Marginalization over $\pi_{ij}$ gives the more traditional point-mixture prior  
\[ \mu_{ij}\sim (1-\frac{a}{a+b}\rho_j)\delta_0+\frac{a}{a+b}\rho_j DP(\gamma N(0,\eta^2)).\] 
Similar structure has been used in the regression context in \cite{lukas06,seo07,carvalho08}. As discussed in those papers, the extended model is able to more adequately shrink towards zero through the induction of zeros for $\pi_{ij}$ and thus can better separate real signals from noise and reduce false discovery. 

Finally, we describe the choice of hyperpriors and the setting of hyperparameters. The base measure of the DP prior for $\mu_j$ is set as a normal distribution with $\mu_0=y_{.j}, \sigma_0^2=\sum_{j=1}^p (y_{.j}-\bar{y})^2/p$ where $y_{.j}=\sum_i y_{ij}/n$ is the observation mean for attribute $j$ and $\bar{y}=\sum_j y_{.j}/n$ is the overall mean of all observations. For the base measure of DP prior for $\sigma_j^2$, we use the vague prior $Inv-Gamma(0.5,0.5)$. Similarly, the standard vague conjugate prior $Inv-Gamma(0.5,0.5)$ is also used as prior for $\eta^2$. For the four concentration parameters in the DPs, $\tau, \alpha, \beta, \gamma$, $Gamma(0.5,0.5)$ is used as the prior. Finally, in the point-mass mixture prior, we follow \cite{lukas06} and set $a=9, b=1, c=0.2, d=199.8$.  

At the end of this section, we emphasize that although our specification of Dirichlet process mixture model has a nested structure in that the cluster component determined by the DP mixture has a base measure on a $p$-dimensional vector whose component has as its distribution a mixture of zero-point mass and a Dirichlet process, the model is entirely different from the so-called nested DP \citep{dunson08}. In a nested DP, the base measure is itself a Dirichlet process. The application in mind for nested DP is clustering of clinic centers with the goal of identifying groups of centers with similar patient outcome distributions. 

\section{Posterior Computation}
Let $c_i, 1\le i\le n$, be the latent class indicator associated with sample $i$, with the specific numbering of no significance, although in the presentation of the algorithm, we assume $1\le c_i\le K$ when $K$ clusters are non-empty. Similarly, $c_j^\mu$ and $c_j^\sigma$ is the cluster indicator for the base-line mean and variance $\mu_j$ and $\sigma_j^2$. We also use $c_{cj}, j=1,\ldots,p,$ to indicate the clustering structure for the $p$ components of the mean vector specific to the $c$-th cluster $\boldmu_c=(\mu_{c1},\ldots,\mu_{cp})$ with $c_{cj}=0$ indicating $\mu_{cj}=0$ is generated from the zero point mass. Similarly as before, we say $c_{ij}=c_{cj}$ if sample $i$ is assigned to cluster $c$. After this augmentation of data, we can update each of the unknowns iterating between the following steps. 

\begin{enumerate}
\item For $j=1,\ldots,p$, draw a new value for $c_j^\mu$ using the following conditional probabilities
\[P(c_j^\mu=c|-)\propto n_{-j,c}^\mu N(\sum_{1\le i\le n}(y_{ij}-\mu_{ij})/n|u,1/v+\sigma_j^2/n),\]
\[P(c_j^\mu\neq c_l^\mu, \forall l\neq j|-)\propto \alpha N(\sum_{1\le i\le n}(y_{ij}-\mu_{ij})/n|\mu_0,\sigma_0^2+\sigma_j^2/n),\]
where $N(x|\mu,\sigma^2)$ denotes the normal density evaluated at $x$, $u=[\mu_0/\sigma_0^2+\sum_{1\le i\le n, k\in C_{-j,c}^\mu}(y_{ik}-\mu_{ik})/\sigma_k^2]/v$, $v=1/\sigma_0^2+\sum_{1\le i\le n, k\in C_{-j,c}^\mu}1/\sigma_k^2$, $C_{-j,c}^\mu$ contains all attribute indices other than $j$ that are assigned to cluster $c$ and $n_{-j,c}^\mu$ is the size of $C_{-j,c}^\mu$.

Then for all $c\in \{c_1^\mu,\ldots,c_p^\mu\}$, draw $\mu_c$ from $N(u,v)$ with $u=[\mu_0/\sigma_0^2+\sum_{1\le i\le n, j\in C_{c}^\mu}(y_{ij}-\mu_{ij})/\sigma_j^2]/v$, $v=1/\sigma_0^2+\sum_{1\le i\le n, j\in C_{c}^\mu}1/\sigma_j^2$, where $C_c^\mu=\{j: 1\le j\le p, c_j^\mu=c\}$.

\item For $j=1,\ldots,p$, draw a new value for $c_j^\sigma$ using the following conditional probabilities
\[P(c_j^\sigma=c|-)\propto n_{-j,c}^\sigma \frac{v^u}{\Gamma(u)}\frac{\Gamma(u+n/2)}{(v+\sum_{1\le i\le n}z_{ij}^2/2)^{u+n/2}},\]
\[P(c_j^\sigma\neq c_l^\sigma, \forall l\neq j|-)\propto \beta \frac{\beta_0^{\alpha_0}}{\Gamma(\alpha_0)}\frac{\Gamma(\alpha_0+n/2)}{(\beta_0+\sum_{1\le i\le n}z_{ij}^2/2)^{\alpha_0+n/2}},\]
where $\Gamma(.)$ is the Gamma function, $z_{ij}=y_{ij}-\mu_j-\mu_{ij}$,  $u=\alpha_0+n_{-j,c}^\sigma\times n/2$, $v=\beta_0+\sum_{1\le i\le n, k\in C_{-j,c}^\sigma}z_{ik}^2/2$, $C_{-j,c}^\sigma$ contains all attribute indices other than $j$ that are assigned to cluster $j$ and $n_{-j,c}^\sigma$ is the size of $C_{-j,c}^\sigma$.

Then for all $c\in \{c_1^\sigma,\ldots,c_p^\sigma\}$, draw $\sigma^2_c$ from $Inv-Gamma(u,v)$ with $u=\alpha_0+n_{c}^\sigma\times n/2$ and $v=\beta_0+\sum_{1\le i\le n, j\in C_{c}^\sigma}z_{ij}^2/2$ where $C_c^\sigma=\{j: 1\le j\le p, c_j^\sigma=c\}$ and $n_c^\sigma$ is its size.

\noindent Now suppose in the current iteration, all non-empty clusters associated with one or more samples are indicated by $\{1,\ldots,K\}$.

\item For $c\in\{1,\ldots,K\}$, $j=1,\ldots,p$, draw $\pi_{cj}$ from the conditional distribution
\[\pi_{cj}|\mu_{cj},\rho_j\sim (1-\rho_j)\delta_0+\rho_j Beta(a,b+1) \mbox{ if } \mu_{cj}=0,\]
\[\pi_{cj}|\mu_{cj},\rho_j\sim Beta(a+1,b) \mbox{ if } \mu_{cj}\neq 0.\]

\item For $j=1,\ldots,p$, draw $\rho_j$ from the conditional distribution
\[\rho_j|\{\pi_{cj}\}_{c=1}^K\sim Beta(c+|\Pi_j|,d+K-|\Pi_j|),\]
where $\Pi_j=\{c: \pi_{cj}>0\}$ is the set of nonzero probabilities $\pi_{cj}$ associated with attribute $j$ and $|\Pi_j|$ is the size of the set.

\item In this step, we update the clustering assignments of the samples $c_i$ as well as the cluster-specific mean vector $\boldmu_{c}$. This basically makes use of Algorithm 7 in \cite{neal00} which is reproduced here for completeness.
 
\begin{enumerate}
\item For $i=1,\ldots,n$, if $c_i$ is not a singleton (i.e. $c_i=c_j$ for some $j\neq i$), let $c_i^*=K+1$ be a new cluster component with $\boldmu_{c_i^*}$ drawn from the prior $\mu_{c_i^*j}\sim (1-a\rho_j/(a+b))\delta_0+a\rho_j/(a+b) DP(\gamma N(0,\eta^2))$. Accept $c_i^*$ with probability
\[\min[1,\frac{\tau}{n-1}\frac{F(\mathbf{y}_i;\boldmu_{c_i^*})}{F(\mathbf{y}_i;\boldmu_{c_i})}],\]

where $F(\mathbf{y}_i;\boldmu_c)=\prod_{j=1}^p N(y_{ij}|\mu_{cj},\sigma_j^2)$. 

If $c_i$ is a singleton, propose $c_i^*=c, c\in\{c_1,\ldots,c_n\}$ with probability $n_{-i,c}/(n-1)$ ($n_{-i,c}$ is the number of samples excluding the $i$-th sample that are currently assigned to cluster $c$) and accept with probability
\[\min[1,\frac{n-1}{\tau}\frac{F(\mathbf{y}_i;\boldmu_{c_i^*})}{F(\mathbf{y}_i;\boldmu_{c_i})}].\]

\item For $i=1,\ldots,n$, if $c_i$ is not a singleton, set $c_i=c, c\in \{c_1,\ldots,c_n\}$ with probability
\[P(c_i=c|-)\propto \frac{n_{-i,c}}{n-1}F(\mathbf{y}_i;\boldmu_c).\]

\item For all $c\in\{c_1,\ldots,c_n\}$, perform a Gibbs sampling step for the components of $\boldmu_{c}|\{\mathbf{y}_i\},\{\rho_j\} \mbox{ with } i\in\{1\le k\le n: c_k=c\}:$ 

\begin{enumerate}
\item For $j=1,\ldots,p$, update $c_{cj}$ from the following distribution
	\begin{eqnarray*}
	P(c_{cj}=0|-)&\propto&(1-\frac{a}{a+b}\rho_j)N(x_j|0,\sigma_j^2/n_c),\\
	\end{eqnarray*}
	for  $c'\in \{c_{ck},1\le k\le p, k\neq j\},$
	\begin{eqnarray*}
	P(c_{cj}=c'|)&\propto&\frac{a}{a+b}\rho_j n_{-j,c'}N(x_j|u,1/v+\sigma_j^2/n_c),\\
	P(c_{cj}\neq c_{ck}\forall k\neq j |)&\propto&\frac{a}{a+b}\rho_j\gamma N(x_j|0,\eta^2+\sigma_j^2/n_c),	
	\end{eqnarray*}
	where $x_j=\sum_{i: c_i=c}(y_{ij}-\mu_j)/n_c$, $u=\sum_{j: c_{cj}=c'}(x_j/\sigma_j^2)/v$ and $v=1/\eta^2+\sum_{j: c_{cj}=c'}1/\sigma_j^2$.
\item For $c'\in \{c_{c1},\ldots,c_{cp}\}$, draw a new value for $\mu_{cc'}$ from $\mu_{cc'}|\{y_{ij}\}, \mbox{ with } c_i=c, c_{cj}=c'$.
\end{enumerate}
\end{enumerate}
Note that here we used a partially collapsed Gibbs step \citep{dyk08} by integrating out $\pi_{ij}$.

\item Draw $\eta^2$ from the conditional distribution
\[\eta^2\sim Inv-Gamma((1+n^\mu)/2, (1+\sum_{(c,j)\in C^\mu}\mu_{cj}^2)/2),\]
where $C^\mu$ is the set of indices of all nonzero unique values of $\mu_{cj}$, and $n^\mu$ is the size of the set.

\item Draw the concentration parameters $\alpha, \beta, \gamma, \tau$ using the data augmentation approach of \cite{escobar95}. 

\end{enumerate}

A note is in order. All the updates above are obtained by standard calculations. In particular, step 5 is a reproduction of Algorithm 7 in \cite{neal00}, with the only difference being step 5(c) where Gibbs sampling must be used since the conditional distribution $\boldmu_{c}$ conditional on samples assigned to cluster $c$ is not directly available in closed form. 

In step 5(a) above, if $c_i$ is a singleton, we draw a new value for $\boldmu_c$ from the prior, which is a high-dimensional vector, and the prior distribution is a mixture of zero point mass and a Dirichlet process. This typically makes $F(\mathbf{y}_i;\boldmu_c)$ extremely small, which is not surprising since $\boldmu_c$ drawn from the prior can hardly explain the observed sample $\mathbf{y}_i$ well. When the update is implemented as presented, this proposal is almost never accepted. To solve this problem, we successfully used a sequential sampling approach that proposes a new value for $\boldmu_c$ taking into account the observed $\mathbf{y}_i$. The proposed sequential sampling step generates the new value with the following scheme:

\begin{itemize}
\item With $c=K+1$, for $j=1,\ldots,p$, draw $c_{cj}$ from the following distribution
	\begin{eqnarray*}
	P(c_{cj}=0|-)&\propto&(1-\frac{a}{a+b}\rho_j)N(x_j|0,\sigma_j^2/n_c),\\
	P(c_{cj}=c'|-)&\propto&\frac{a}{a+b}\rho_j n_{-j,c}N(x_j|u,1/v+\sigma_j^2/n_c),\; c'\in \{c_{ck}:k\neq j\},\\
	P(c_{cj}\neq c_{ck} \forall k\neq j |-)&\propto&\frac{a}{a+b}\rho_j\alpha N(x_j|0,\eta^2+\sigma_j^2/n_c),	
	\end{eqnarray*}
	where we conditioned on $\{\mu_{ck}\}_{k=1}^{j-1}, x_j, \{\sigma_k\}_{k=1}^p,\rho_j$ and $\eta$. In the above we define $x_j=\sum_{i: c_i=c}(y_{ij}-\mu_j)/n_c$, $u=\sum_{j: c_{cj}=c'}x_j/\sigma_j^2/v$ and $v=1/\eta^2+\sum_{j: c_{cj}=c'}1/\sigma_j^2$. 
\item For $c'\in \{c_{k1},\ldots,c_{kp}\}$, draw a new value for $\mu_{cc'}$.
\end{itemize}
These expressions are very similar to the Gibbs step 5(c), with the important difference that when proposing new value for $c_{cj}$, only the previously sampled $c_{ck}, k<j$ are available and the update is performed sequentially. With this change of proposal distribution, the acceptance probability in step 5(a) should be changed to
\[\min[1,\frac{\tau}{n-1}\frac{F(\mathbf{y}_i;\boldmu_{c_i^*})}{F(\mathbf{y}_i;\boldmu_{c_i})}\frac{Q_0(\boldmu_{c_i^*})}{Q(\boldmu_{c_i^*})}]\]
and
\[\min[1,\frac{n-1}{\tau}\frac{F(\mathbf{y}_i;\boldmu_{c_i^*})}{F(\mathbf{y}_i;\boldmu_{c_i})}\frac{Q(\boldmu_{c_i})}{Q_0(\boldmu_{c_i})}]\]
respectively, where $Q_0$ is the proposal density when $\boldmu_c$ is drawn from the prior and $Q$ is the proposal density of $\boldmu_c$ when it is drawn from the sequential sampling approach described above.

Finally, for inferences on cluster-specific parameters, such as $\mu_{cj}$ and $\pi_{cj}$, we need to take care of the label switching problem \citep{stephens00,tadesse05}. This can be done conditionally on the number of clusters for the samples, $K$. We refer the readers to \cite{tadesse05} for details.

\section{Extension to Variance Shifts}
Although it is not our focus in this paper, our model can easily be extended to the case where groups are distinguished by both different mean and variance for one or more attributes. On can extend the model and write the observed data as
\[y_{ij}=\mu_j+\mu_{ij}+\sigma_j\sigma_{ij}\epsilon_{ij},\;\epsilon_{ij}\stackrel{i.i.d.}{\sim}N(0,1).\]
The extra factor $\sigma_{ij}$ indicates the difference in variance for data assigned to different groups. 
Similarly to the mean shift model, we can put a hierarchical sparsity prior 
\begin{eqnarray*}
\sigma_{ij}^2&\sim& (1-\pi_{ij}^\sigma)\delta_1+\pi_{ij}^\sigma DP(\kappa Inv-Gamma(\alpha_\sigma,\beta_\sigma)),\\
\pi_{ij}^\sigma&\sim& (1-\rho_j^\sigma)\delta_0+\rho_j^\sigma Beta(a,b),\\
\rho_j^\sigma&\sim&Beta(c,d).
\end{eqnarray*}

Note that for variances, sparsity means many of the $\sigma_{ij}$ will be exactly equal to one. For the base measure, we should also choose it to be roughly centered at one for identifiability, for example, we can use $\alpha_\sigma=1.5, \beta_\sigma=0.5$ as a vague prior.

This extension causes few extra complications on the updating strategy for posterior computation, with extra updates for $\sigma_{ij}$ as well as some slight change of formula. The details are omitted here. We do not consider further the case with variance shift since one can argue that for the microarray analysis for example, the researchers typically focus on mean shift as a distinguishing feature of tissue subtypes.

\section{Simulation and Application}
\subsection{Simulation Study}
\begin{figure}
\centerline{
\subfigure[]{\includegraphics[width=2.5in]{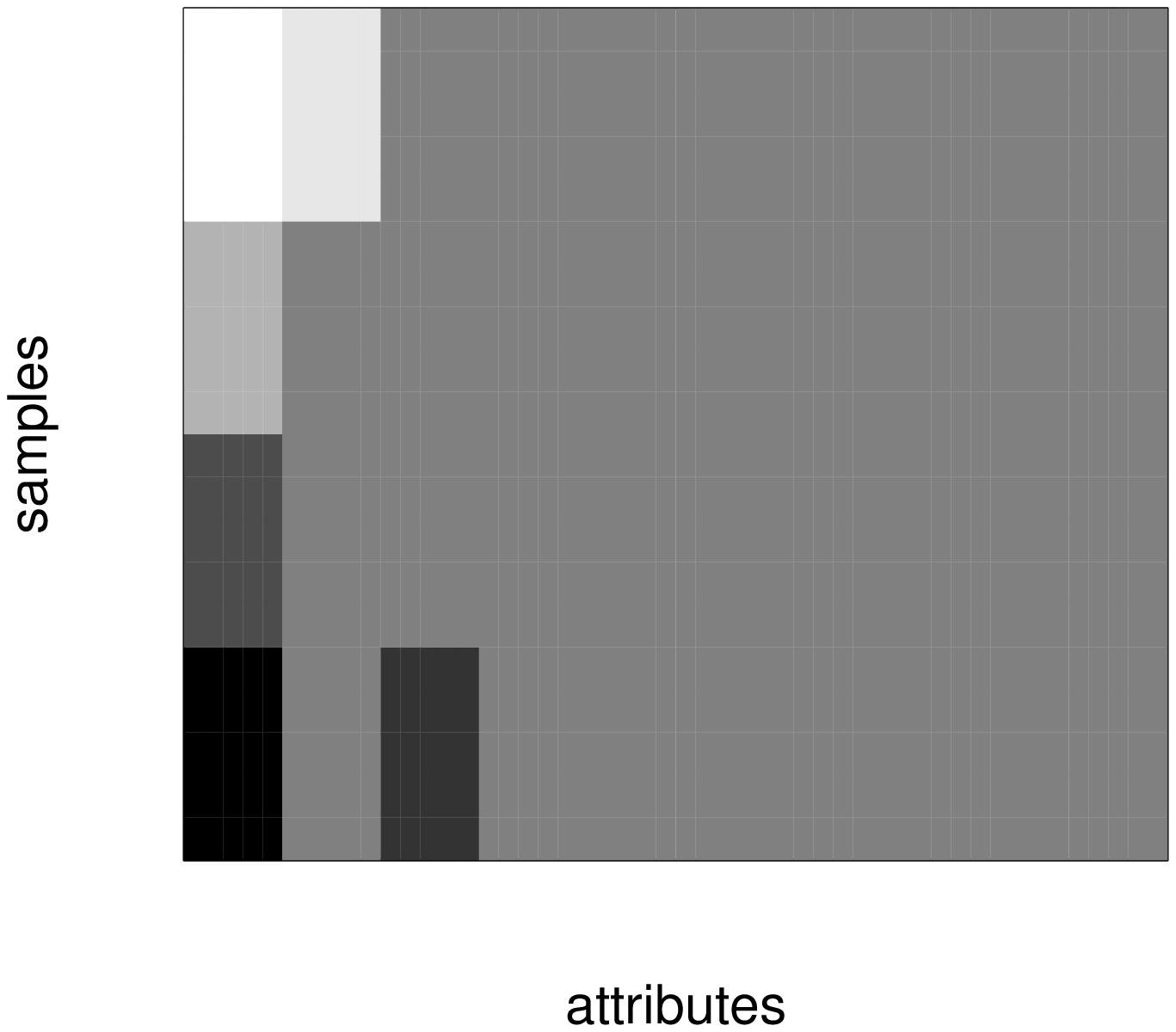}
}
}

\centerline{\subfigure[]{\includegraphics[width=2.5in]{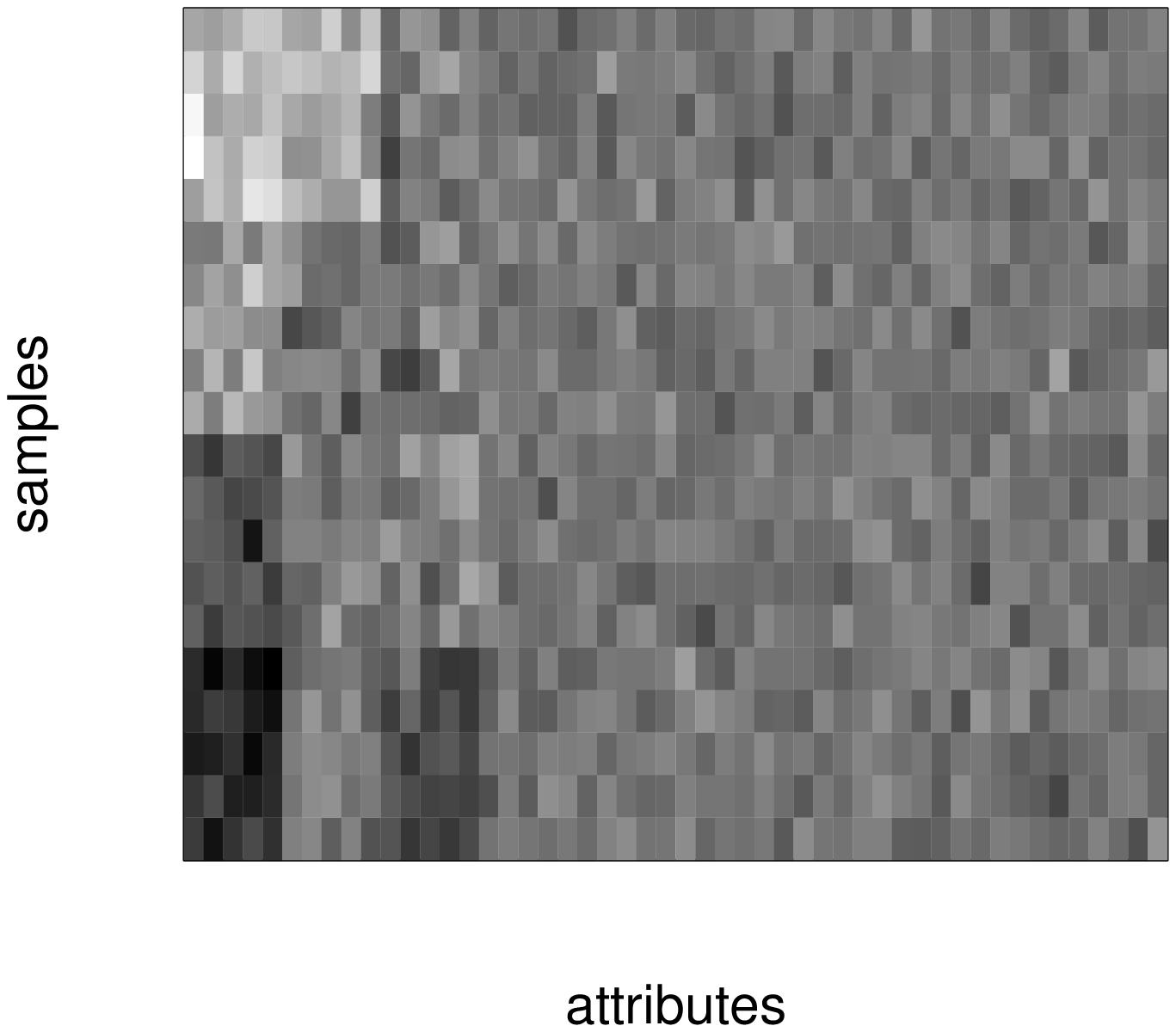}
}
\hfil
\subfigure[]{\includegraphics[width=2.5in]{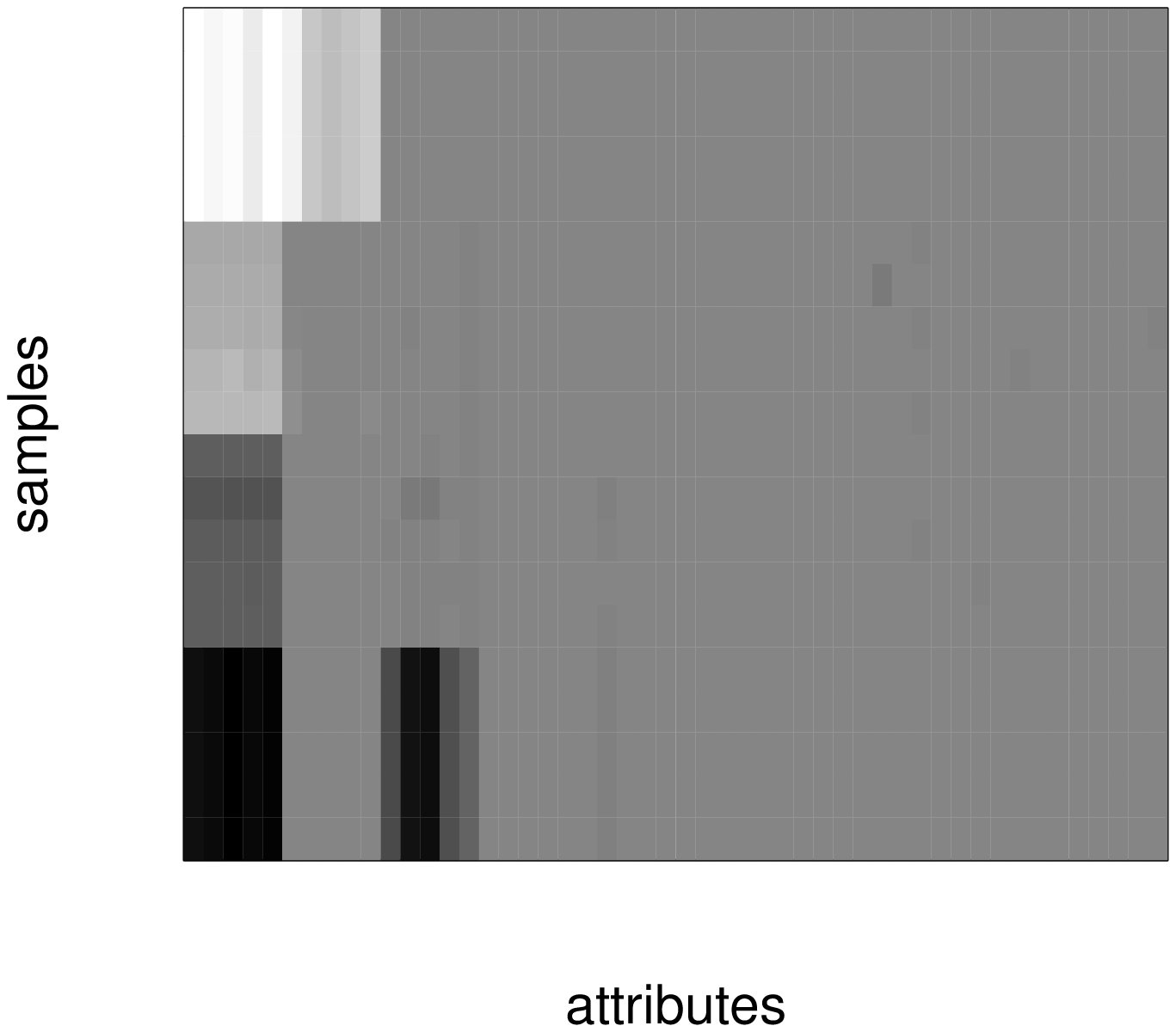}
}}

\caption{(a) Mean structure $\mu_{ij}$ for the simulated data plotted as an image. (b) Noisy observed data. (c) Estimated mean under our approach.\label{Y}   }
\end{figure}

\begin{figure}
\includegraphics[width=6in]{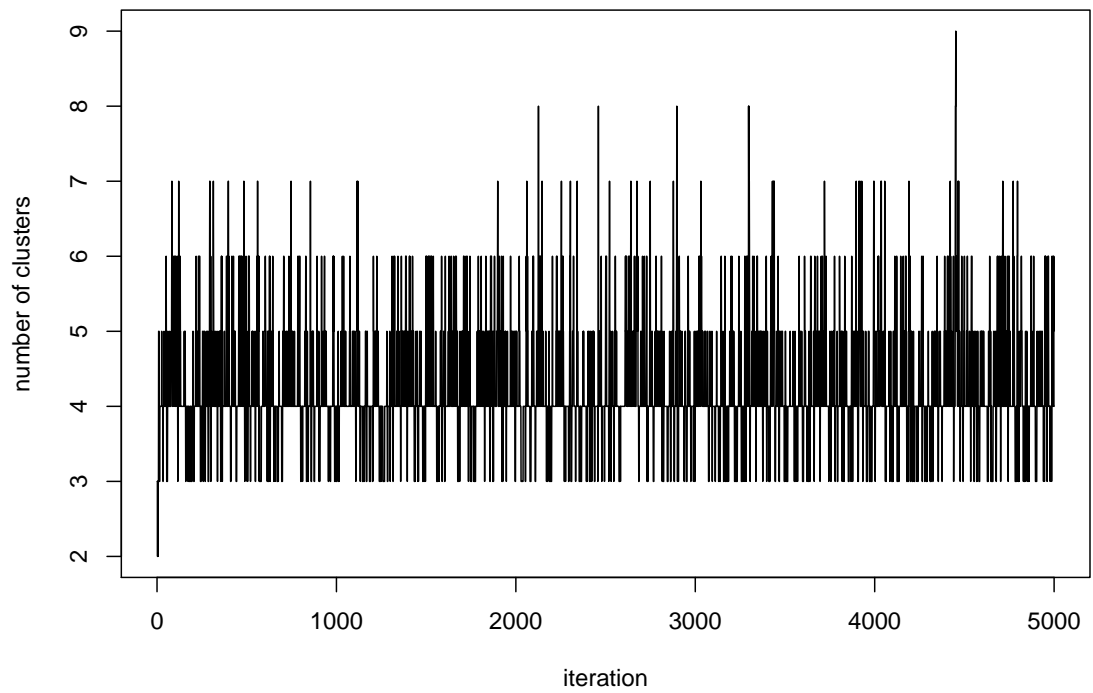}
\caption{Trace plot of the number of clusters $K$ for the simulated dataset.\label{simulate.K.convergence}}
\end{figure}

\begin{figure}
\centerline{
\subfigure[]{\includegraphics[width=2.5in]{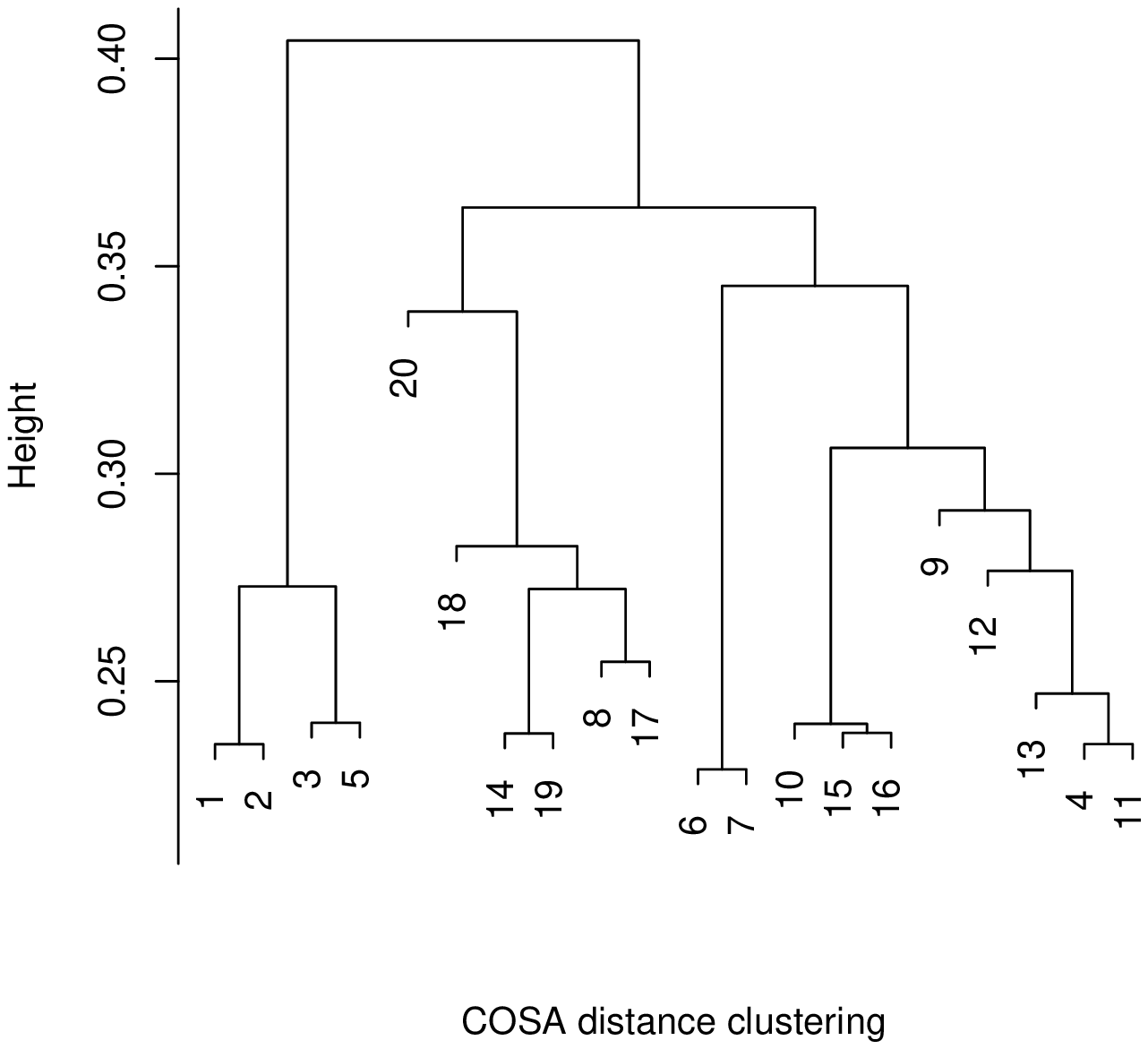}
}
}

\centerline{\subfigure[]{\includegraphics[width=2.5in]{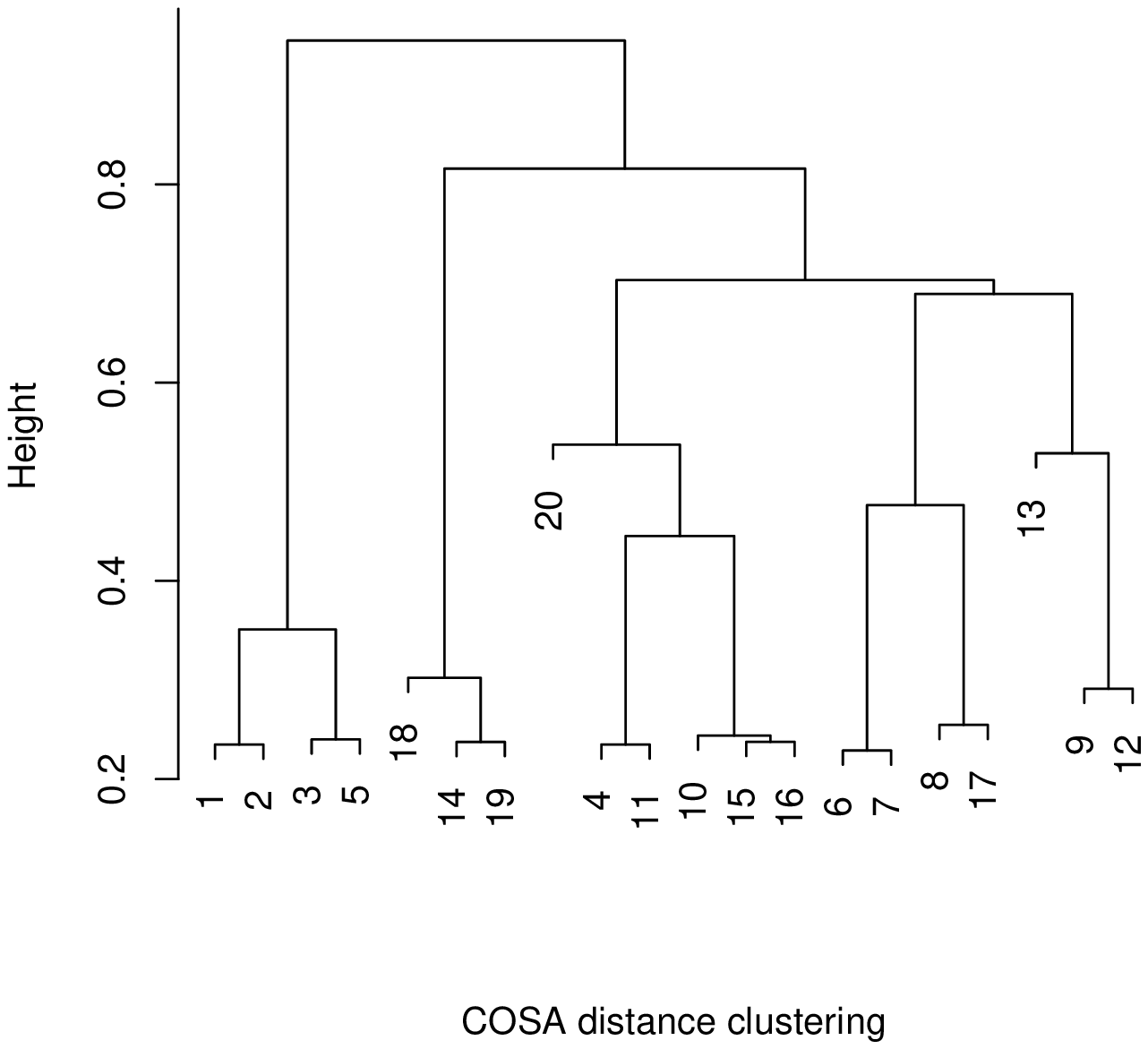}
}
\hfil
\subfigure[]{\includegraphics[width=2.5in]{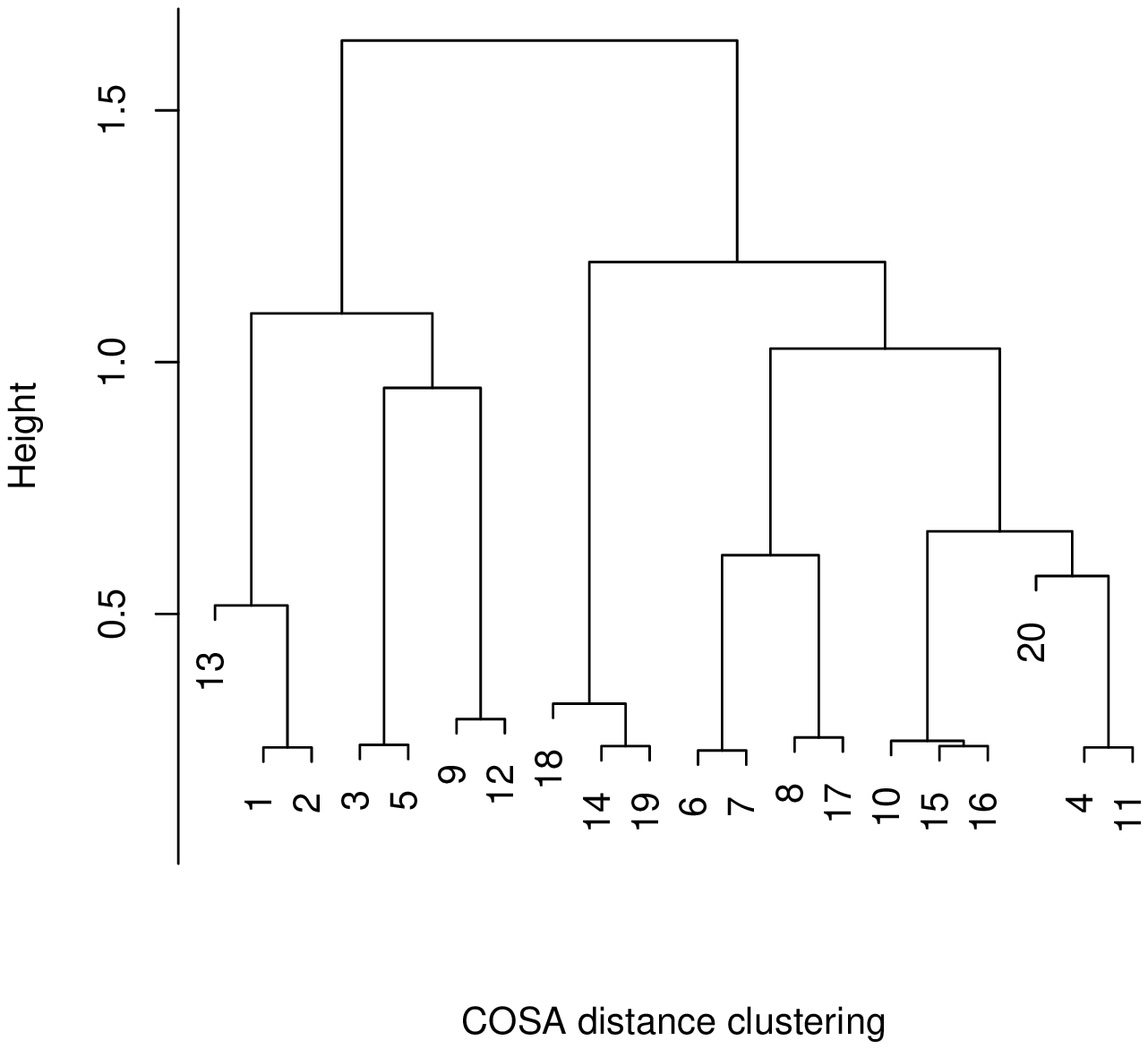}
}}

\caption{Results from hierarchical clustering with COSA algorithm based on (a) single linkage; (b) average linkage; (c) complete linkage.\label{cosa}   }
\end{figure}

We investigate the performance of our estimation method in a simulation study. A dataset containing 20 samples and 200 covariates is generated as follows.
\begin{eqnarray}
y_{ij}&=&\mu_{ij}+\sigma_j\epsilon_{ij}, \; \epsilon_{ij}\sim N(0,1),\nonumber\\
\mu_{i,j}&=&0.25, 1\le i\le 5, 1\le j\le 5,\nonumber\\
\mu_{i,j}&=&0.1, 6\le i\le 10, 1\le j\le 5,\nonumber\\
\mu_{i,j}&=&-0.1, 11\le i\le 15, 1\le j\le 5,\nonumber\\
\mu_{i,j}&=&-0.25, 16\le i\le 20, 1\le j\le 5,\nonumber\\
\mu_{i,j}&=&0.2, 1\le i\le 5, 6\le j\le 10,\nonumber\\
\mu_{i,j}&=&-0.15, 16\le i\le 20, 11\le j\le 15,\nonumber\\
\mu_{ij}&=& 0 \mbox{ otherwise},\nonumber\\
\sigma_j&=&0.1, 1\le j\le 15,\nonumber\\
\sigma_j&=&0.05, \mbox{ otherwise}.\nonumber
\end{eqnarray}
The structure of $\mu_{ij}$ is shown in Figure \ref{Y}(a) where different values for $\mu_{ij}$ show up as different gray levels. Each row in the image represents a sample and each column represents an attribute. Only the first 50 attributes are shown. The first 5 covariates distinguishes across all four groups, while attributes 6-10 distinguish the first cluster from the others and attributes 11-15 distinguish the fourth cluster from the others. We use the model described in Section 2 to fit the simulated dataset. Figure \ref{Y}(b) shows the observed data in the same format as Figure \ref{Y}(a). Figure \ref{simulate.K.convergence} shows the posterior updates for the number of clusters identified as well as gives some indication of the mixing of the Markov chain. The posterior gives strong support for four clusters, with support for five clusters comes next. In simulation as well as real data application that follows, we used a burn-in period of $10,000$ updates and $40,000$ iterations after burn-in for inferences. 
The posterior estimates of $\mu_{ij}$ is shown in Figure \ref{Y}(c) as a matrix for the first $50$ attributes only. Four clusters and the zero structures are clearly identified. In contrast, the hierarchical clustering based on COSA algorithm of \cite{friedmanjrss04} failed to identify the true clusters, as shown in Figure \ref{cosa} for single, average, and complete linkage.

\begin{figure}
\includegraphics{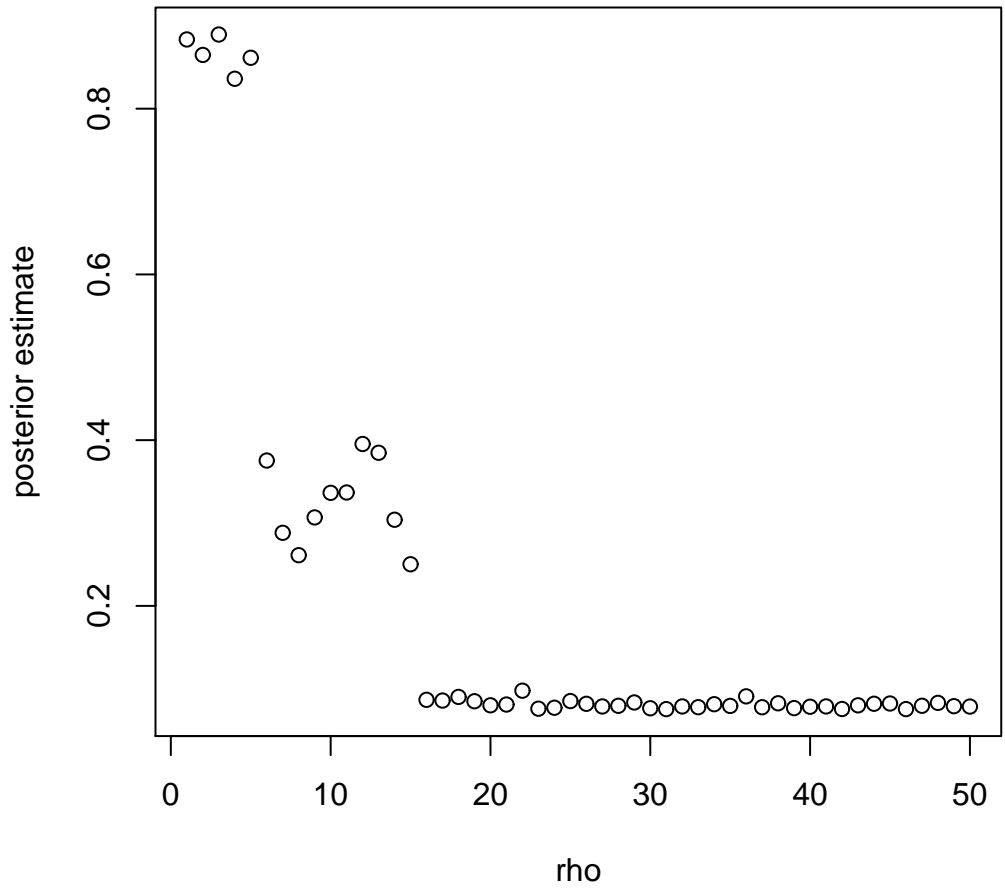}
\caption{Posterior estimates of $\rho_j$ for $1\le j\le 50$.\label{rho}}
\end{figure}

In Figure \ref{rho}, we show the posterior estimate of $\rho_j$ for $1\le j\le 50$, which indicates the contribution of the $j$-th attribute to cluster discrimination. The results are quite encouraging, with the first 15 attributes clearly identified as signal variables and the first 5 attributes estimated to be associated with larger values of $\rho_j$, consistent with the simulation scheme. We can also use a simple threshold of $0.5$ on posterior estimates of $\pi_{ij}$. In particular, we decide attribute $j$ to be relevant for clustering if $\pi_{ij}>0.5$ for at least one $i$. This strategy also exactly identifies the first $15$ attributes as significant.  

Finally, for this simulated example, using DP for $\mu_j$ and $\mu_{ij}, 1\le j\le p$ performs better than a normal prior (corresponding to the case with $\alpha\rightarrow\infty$ and $\gamma\rightarrow\infty$. The mean squared error of $\mu_j+\mu_{ij}, 1\le i\le 20, 1\le j\le 15$, under our model is $0.006$, in contrast with $0.011$ when $\alpha,\gamma\rightarrow\infty$. This is consistent with the results reported in \cite{nott08}. 

Now we use additional simulated examples with various choices of the number of attributes $p$ and the values for mean vectors to investigate the performance of our method. Besides the example above, we use the following simulation schemes. All examples are simulated from model $y_{ij}=\mu_{ij}+\sigma_j\epsilon_{ij}$ as before.

\begin{itemize}
\item Example 2. Same as the example presented above, except the number of noise variables are increased to $p=1000$.
\item Example 3. Here we have $n=20$ samples and $p=50$ attributes, among which $10$ attributes are informative for clustering across all four clusters. The cluster sizes are chosen to be $3,3,7,7$ respectively, with $\mu_{ij}=c/4$ if sample $i$ belongs to cluster $c$, $1\le c\le 4$, when $1\le j\le 10$, and $\sigma_j=0.1$ for all $j$.
\item Example 4. Here $n=20$ and $p=50$, with $\mu_{ij}=j/50$ when $1\le i\le 10$, $\mu_{ij}=(50-j)/50$ when $11\le i\le 20$, and $\sigma_j=0.1$ for all $j$. 
\end{itemize}
Our method is compared to two other model-based Bayesian clustering methods with variable selection proposed in \cite{hoff06} and \cite{kim06}. For each example, we simulated $50$ datasets. The different methods are compared using three performance measures. First, we compute the average mean squared errors for $\mu_{ij}$ where we only consider attributes that are relevant for clustering. Also, we consider the number of attributes selected by each method as well as its overlap with true relevant attributes. Based on the existing implementation for Hoff's approach, the attributes are selected such that they maximize the joint posterior distribution. Besides, all three approaches can identify the correct number of clusters in all examples.

The results in Table \ref{tab:sim} show that our approach always outperformed the other two for all the examples considered here in terms of mean squared error. Examples 1 through 3 are perhaps favorable to our approach. Example 4 was intended for a situation where all components for a cluster mean are distinct. Our method still works well in this situation in terms of mean squared error, compared with the other two approaches. In terms of selected variables, the approach of \cite{hoff06} tends to select a large number of relevant variables resulting in high false discovery rate. When there exist attributes that only distinguish a subset of cluster components, such as the situation in Examples 1 and 2, the variable selection approach of \cite{kim06} tends to miss some of the those covariates. Note that the approach of \cite{kim06} cannot give informations on whether an attribute only distinguishes a subset of the cluster components even if the attribute is selected.

\subsection{Leukemia Gene Expression Data Example}
We use the leukemia gene expression dataset \citep{golub99} to demonstrate the utility of our proposed method. The training dataset contains 38 tissue samples, among which 11 samples are acute myeloid leukemia (AML) and the rest are acute lymphoblastic leukemia (ALL). The 27 ALL samples are further divided into two subgroups: 8 T-cell and 19 B-cell samples. The samples were arrayed with a total of 7129 genes in a microarray experiment. Following the standard preprocessing steps in \cite{dudoit02}, we truncate the expression values to within the interval $[1,16000]$, and delete those genes whose maximum and minimum expression across all samples satisfies $\max/\min\le 5$ and $\max-\min\le 500$. Finally, we select the top $2000$ genes with the largest variances across all samples so that at the end we have for this dataset $n=38, p=2000$. 

We apply our proposed method to the dataset with the hyperparameters set exactly as discussed in Section 2. Convergence of the MCMC updates is invariably a concern in high-dimensional problem with variable selection. As a simple diagnostic, two MCMC runs of 50,000 iterations with the first 10,000 as burn-in are implemented, with different initialization. In particular, we start one Markov chain with initially all samples assigned to one cluster, and another chain where each sample is assigned to its own separate cluster. The posterior estimates of various unknown quantities for the two runs shows good agreement which indicates the chains mixed well in our implementation.

\begin{figure}
\includegraphics{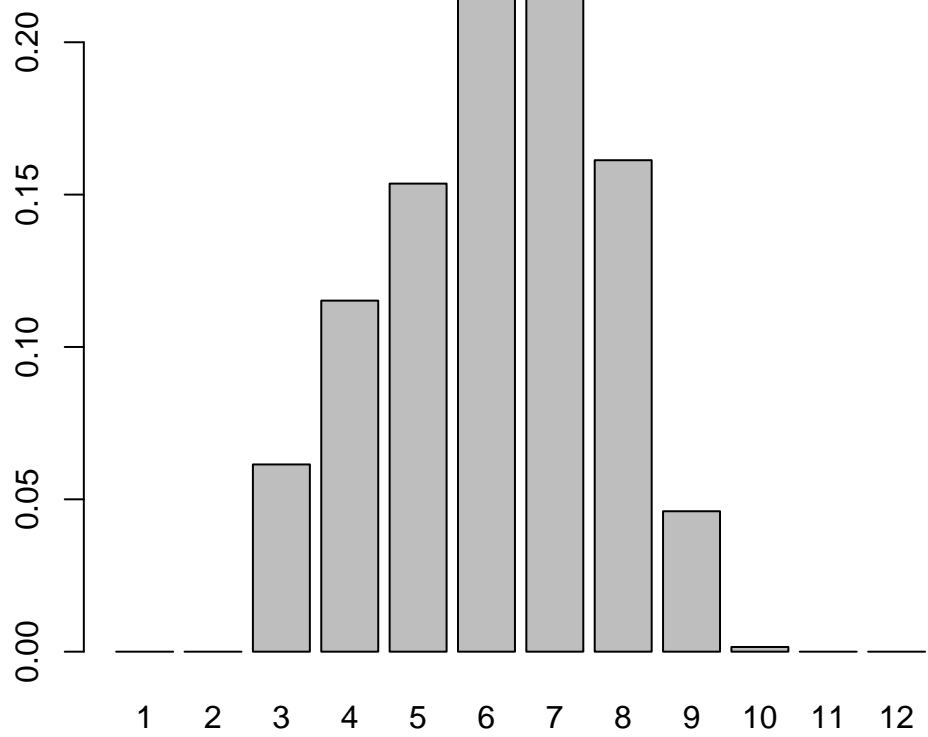}
\caption{Posterior estimate of the number of clusters for the leukemia expression dataset.\label{golubnumbercluster}}
\end{figure}

As shown in Figure \ref{golubnumbercluster}, the posterior for this dataset put most of the support for the number of clusters between 3 and 9, with 6 clusters receiving the highest score. Conditional on $K=6$, setting the threshold $0.5$ for the posterior estimates of $\pi_{cj}$ returns $872$ genes. This is much larger than the $120$ genes reported in \cite{kim06}. Previous studies, such as \cite{thomas01}, also demonstrated that there were a large number of genes differentially expressed between different tissue samples.  

For inference about the cluster structure and comparison to known tissue subtypes, we estimate the posterior probability of $c_i=c, 1\le c\le 6$ from posterior samples conditioned on $K=6$ with the help of the procedure that deals with label switching. Each sample is allocated to the cluster with the largest posterior probability.  The relationship between this allocation and known tissue types are shown in Table \ref{golubtable}. Under our method, one of the ALL samples is misclassified into a cluster dominated by AML samples, also one of the AML samples is misclassified into a cluster dominated by ALL samples. We see that the known AML and ALL-B tissue types might further consist of some subtypes. Using our method, we can also discover genes that distinguish only some subgroups. For example, among those $872$ genes relevant for clustering only $64$ of them can distinguish between ALL and AML samples without discriminative power for different subtypes.

\section{Conclusion}
In this article, we propose a novel Bayesian approach to high-dimensional clustering with variable selection. The distinguishing features of our method include a separate Dirichlet process for shrinkage estimation of cluster mean, as well as a hierarchical point-mass structure that improves the separation of significant signal from noise variables. We propose a sequential sampling approach in one of the updating iterations of the MCMC algorithm to solve the computational problem associated with the high dimensionality of the mean vector. 

Our approach only involves diagonal covariance matrices. It has been argued in other studies in both supervised and unsupervised context that for ``high dimensional low sample size" setting, this working independence assumption is more effective than the full covariance matrix approach \citep{fraley06,bickel04,tibshirani03}. Generalization of our method to general covariance structure seems much more challenging. 

In our implementation, we choose to use Algorithm 7 in \cite{neal00} for its simplicity in implementation. More efficient approaches like split-merge update \citep{jain04} can also be utilized. Nevertheless, we still expect the original algorithm should be modified using sequential sampling instead of drawing new components from the prior to make the implementation feasible in practice. 

\bibliographystyle{jasa}
\bibliography{papers,books,DPpaper}

\newpage

\begin{table}
\caption{Simulation results for the four examples. The numbers in each cell correspond to median mean squared error, number of attributes selected and overlap between selected attributes and the truth, respectively.\label{tab:sim}}
\vspace{0.1in}
{
\begin{tabular}{c|c|c|c|c}
\hline\hline
Methods & Example 1& Example 2& Example 3& Example 4\\\hline
Our approach & 0.005 15 15 &0.021 18 13&0.012 \phantom{0}8 \phantom{0}8&0.016 46 46 \\
\cite{hoff06}& 0.013 19 15 &0.046 26 10&0.027 17 \phantom{0}7&0.025 50 49\\
\cite{kim06} & 0.018 14 14  &0.069 14 \phantom{0}8&0.016 11 10&0.021 48 48  \\
\end{tabular}}
\end{table}

\begin{table}
\caption{Clustering results for leukemia expression data conditional on $K=6$.\label{golubtable}}
\vspace{0.1in}
{\renewcommand{\arraystretch}{1.5}
\renewcommand{\tabcolsep}{0.5cm}
\begin{tabular}{c|c|c|c|c|c|c}
\hline\hline
 & \multicolumn{6}{|c}{cluster from the proposed method}\\
samples& 1&2&3&4&5&6\\\hline
ALL-T(8) & 0&0 &8 &0 &0 &0\\
ALL-B(19)&0 &1 &0 &6 &4 &8  \\
AML(11)  &7 &3  &0&0 &1 &0  \\
\end{tabular}}
\end{table}

\end{document}